\documentstyle[12pt,aaspp4]{article}

\begin{document}

\title{DARK MATTER IS BARYONS} 

\author{Robert K. Soberman}
\affil{2056 Appletree Street, Philadelphia, PA 19103}
\authoremail{soberman@sas.upenn.edu}
\author{Maurice Dubin}
\affil{14720 Silverstone Drive, Silver Spring, MD 20905}
\authoremail{mdubin@aol.com}

\lefthead{Soberman \& Dubin} 
\righthead{DARK MATTER IS BARYONS}
 
\begin{abstract}  

A comet-like, but magnitudes smaller, extremely low albedo interstellar meteoroid population of fragile aggregates with solar type composition, measured in space and terrestrially, is most probably the universal dark matter. Although non-baryonic particles cannot be excluded, only "Big Bang" cosmology predicts an appreciable fraction of such alternate forms. As more counter-physics hypotheses are added to fit observation to the expanding universe assumption, a classical physics alternative proffers dark matter interactive red shifts normally correlated with distance.  The cosmic microwave background results from size-independent thermal plateau radiation that emanates from dark matter gravitationally drawn into the Galaxy.

\end{abstract}


\section{Cosmoids}

The existence of a meteoroid population in highly eccentric orbits was discovered in the results of the three "dust" experiments carried aboard the PIONEER 10 and 11 missions to the outer planets. A re-examination of the results from the Asteroid/Meteoroid Detector (nicknamed Sisyphus) established beyond statistical doubt that a new population termed cosmoids, a contraction of cosmic meteoroids, had been measured \markcite{Dubin91} (Dubin \& Soberman 1991). The three-way conflict among the conclusions of the three meteoric experiments \markcite{Soberman76} (Soberman et al.\null\ 1976) was resolved by the recognition that all had measured cosmoids \markcite{Dubin91} (Dubin \& Soberman 1991). 

The seminal velocity measurements of Gr\"un et al.\null\ \markcite{Gruen92} (1992) established what was previously suspected, that cosmoids were extra-solar in origin. That was confirmed when velocities in excess of 100 km/s were measured for members of a ``new population'' of terrestrial radio meteors \markcite{Taylor96} (Taylor et al.\null\ 1996) first discovered utilizing VLF (2 MHz) radar \markcite{Olsson87} (Olsson-Steel \& Elford 1987).

The properties that allowed cosmoids to escape detection through more than a half century of intensive search for their predicted existence includes their extremely low comet-nucleus-like albedo (about 2 - 3

\section{Are Non-baryons Permitted?}

An appreciable fraction of non-baryonic dark matter cannot be excluded by the discovery of cosmoids.  However, only ``Big Bang'' theory predicts sizable masses in alternate forms.  Apart from detected MACHOS (MAssive Compact Halo Objects) which are likely large gravity compressed cosmoids, the nature of other hypothesized candidates might have been drawn from ``Alice In Wonderland'' \markcite{Cowen00} (Cowen 2000). Occam's razor argues that they should not be assumed if unneeded.

The validity of the ``Big Bang'' has been long questioned by a small distinguished group of astrophysicists beginning with Fritz Zwicky.  He coined its name to deride the concept.  For three and one-half decades Halton Arp \markcite{Arp66} (1966) has presented hundreds of examples where the red shift conflicts with apparently associated objects. Only Penzias' and Wilson's \markcite{Penzias65} (1965) discovery of the approximate 3 K Cosmic Microwave Background (CMB) shifted most to believe in an expanding universe and fostered a major cosmology industry.  Neglected or dismissed is the need for ``inflation,'' faster-than-light galactic aggregation and most recently the resurrection of Einstein's cosmological constant investing free space with anti-gravity power. 

An alternate explanation for the red shift that ties it generally with distance is the interaction of photons with cosmoids \markcite{Soberman01} (Soberman \& Dubin 2001). In this model the CMB results from a thermal (size-independent) eigenvalue plateau for cosmoids being drawn into the Galaxy.  The fractional kelvin discrepancy may be removed by laboratory determination.  This contrasts with the decade discrepancy between the measured CMB and prediction \markcite{Alpher48} (Alpher et al.\null\ 1948).

\end{document}